\documentclass[12pt]{iopart}
\usepackage{graphicx, amssymb, amsfonts,braket}

\newcommand{\be}[0]{\begin{equation}}
\newcommand{\ee}[0]{\end{equation}}
\newcommand{\bea}[0]{\begin{eqnarray}}
\newcommand{\eea}[0]{\end{eqnarray}}

\begin{document}

\title{High-dimensional quantum cryptography with twisted light}
\author{M~Mirhosseini$^1$, O~S~Maga\~na-Loaiza,$^1$, M N O'Sullivan$^1$, \mbox{B Rodenburg$^1$}, M~Malik$^{1,2}$, M P J Lavery$^3$, M J Padgett$^3$, \mbox{D J Gauthier}$^4$, and R~W~Boyd$^{1,5}$}
\address{$^1$ The Institute of Optics, University of Rochester, Rochester, New York 14627}
\address{$^2$ Institute for Quantum Optics and Quantum Information (IQOQI), Austrian Academy of Sciences, Boltzmanngasse 3, A-1090 Vienna, Austria}
\address{$^3$ School of Physics and Astronomy, University of Glasgow, Glasgow, G12 8QQ, UK}
\address{$^4$ Department of Physics, Duke University, Durham, NC 27708 USA}
\address{$^5$ Department of Physics, University of Ottawa, Ottawa ON K1N 6N5, Canada}
\ead{mirhosse@optics.rochester.edu}

\begin{abstract}
Quantum key distributions (QKD) systems often rely on polarization of light for encoding, thus limiting the amount of information that can be sent per photon and placing tight bounds on the error that such a system can tolerate. Here we describe a proof-of-principle experiment that indicates the feasibility of high-dimensional QKD based on the transverse structure of the light field, allowing for the transfer of more than 1 bit per photon. Our implementation uses the orbital angular momentum (OAM) of photons and the corresponding
  mutually unbiased basis of angular position (ANG). Our experiment uses a digital
  micro-mirror device for the rapid generation of OAM and ANG modes at 4 kHz, and a
  mode sorter capable of sorting single photons based on their OAM and
  ANG content with a separation efficiency of 93\%. Through the use of a
  7-dimensional alphabet encoded in the OAM and ANG bases, we achieve a channel
  capacity of 2.05 bits per sifted photon. Our experiment shows that, in addition to having an increased
  information capacity, QKD systems based on spatial-mode encoding will be more tolerant to
  errors and thus more robust against eavesdropping attacks.
  \end{abstract}


\maketitle

\section{Introduction}

First introduced in 1984 by Bennett and Brassard, quantum key distribution (QKD) is a method for distributing a secret key between two parties \cite{Bennett1984,Gisin:2002gb}. Due to a fundamental property of quantum physics known as the no-cloning theorem, any attempt made by a third party to eavesdrop inevitably leads to errors that can be detected by the sender and receiver \cite{Wootters:1982ex}. Modern QKD schemes conventionally use a qubit system for encoding information, such as the polarization of a photon. Such systems are easily implemented because technology for encoding and decoding information in a qubit state-space is readily available today. Recently, the spatial degree of freedom of photons has been identified as an extremely useful resource for transferring information \cite{MolinaTerriza:2007ig,Gibson2004}. The information transfer capacity of classical communication links has been increased to more than one terabit per second using spatial-mode multiplexing \cite{Wang2012}. In addition, it has been shown that employing multilevel quantum states (qudits) can increase the robustness of a QKD system against eavesdropping \cite{Cerf:2002fp,AliKhan:2007ia,Zhang:2014gt}. Considering the above benefits, it is expected that realizing a large alphabet by means of spatial-mode encoding can drastically enhance the performance of a QKD system.

The feasibility of high-dimensional QKD in the spatial domain has been previously demonstrated by encoding information in the transverse linear momentum and position bases \cite{Walborn:2006jv,Etcheverry:2013fu}. 
While such encoding schemes provide a simple solution for increasing the information capacity, they are not suitable for long-haul optical links due to the cross-talk caused by diffraction.  Diffraction creates a varying transmission loss for different spatial frequencies that results in mixing of spatial modes \cite{Shapiro1974,Miller:2000tw}. Cross-talk increases the quantum bit error rate (QBER) and fundamentally limits the secure key rate of a QKD system. This adverse effect can be alleviated by employing OAM modes. Due to their rotational symmetry, OAM modes have the desirable property of remaining orthogonal upon propagation in a system with round apertures \cite{Slepian:1965wz}. In a recent experiment, OAM modes were used for performing classical communication over a 3-km-long free-space optical link \cite{Krenn:2014wj}. In this experiment, the main contribution of turbulence was shown to be a transverse displacement of the OAM modes \emph{i.e.}, a tip-tilt error that  can be readily corrected for with present technology. As such, OAM modes are ideal candidates for use in a free-space quantum communication link.

The OAM of photons has enabled many applications in the field of quantum information including studies of quantum entanglement \cite{Mair:2001fd,Leach:2010bm,McLaren:2014ec,Leach:2012wj}, photonic superdense coding \cite{Barreiro:2008jl}, realization of high-dimensional wavefunctions \cite{Malik:2014bf}, and quantum cloning \cite{Nagali:2009hy}. Similarly, a number of studies have investigated the benefits of employing OAM modes in quantum cryptography \cite{Mafu:2013vr,Groblacher2006,Vaziri:2002fx}. Recently, rotation-invariant OAM vector modes have been used for performing QKD in a two-dimensional state space \cite{Vallone:2014vc}. Although this method offers an advantage in terms of optical alignment, it fails to utilize the large information capacity of the OAM basis. The complete realization of a high-dimensional QKD system with OAM has remained impractical up until now mostly due to the difficulty in efficiently sorting single photons in the OAM basis. Additionally, any realistic application requires a fast key-generation rate that cannot be achieved with most of the common methods for generating OAM modes.

\section{High-dimensional QKD with orbital angular momentum}

In this paper we describe a proof-of-principle experiment that demonstrates the feasibility of performing high-dimensional quantum cryptography with OAM modes. We encode information in a 7-dimensional set of OAM modes along with modes in the mutually unbiased basis of azimuthal angle (ANG). Our scheme uses a digital micro-mirror device (DMD) for fast generation, and an efficient technique for unambiguous sorting of both OAM and ANG modes. By combining these techniques, we selectively generate the set of 14 spatial modes at a rate of 4 kHz and correspondingly detect them with a separation efficiency of 93$\%$. We measure a channel capacity of 2.05 bits per sifted photon with an average symbol error rate of 10.5$\%$ that is well below the error bounds that are required for security against intercept-resend eavesdropping attacks. Of our error rate, 4\% is due to detector dark counts and ambient light, and 6.5\% is due to errors due to intermodal crosstalk.

\begin{figure}[t]
\centerline{\includegraphics[width =0.8 \textwidth]{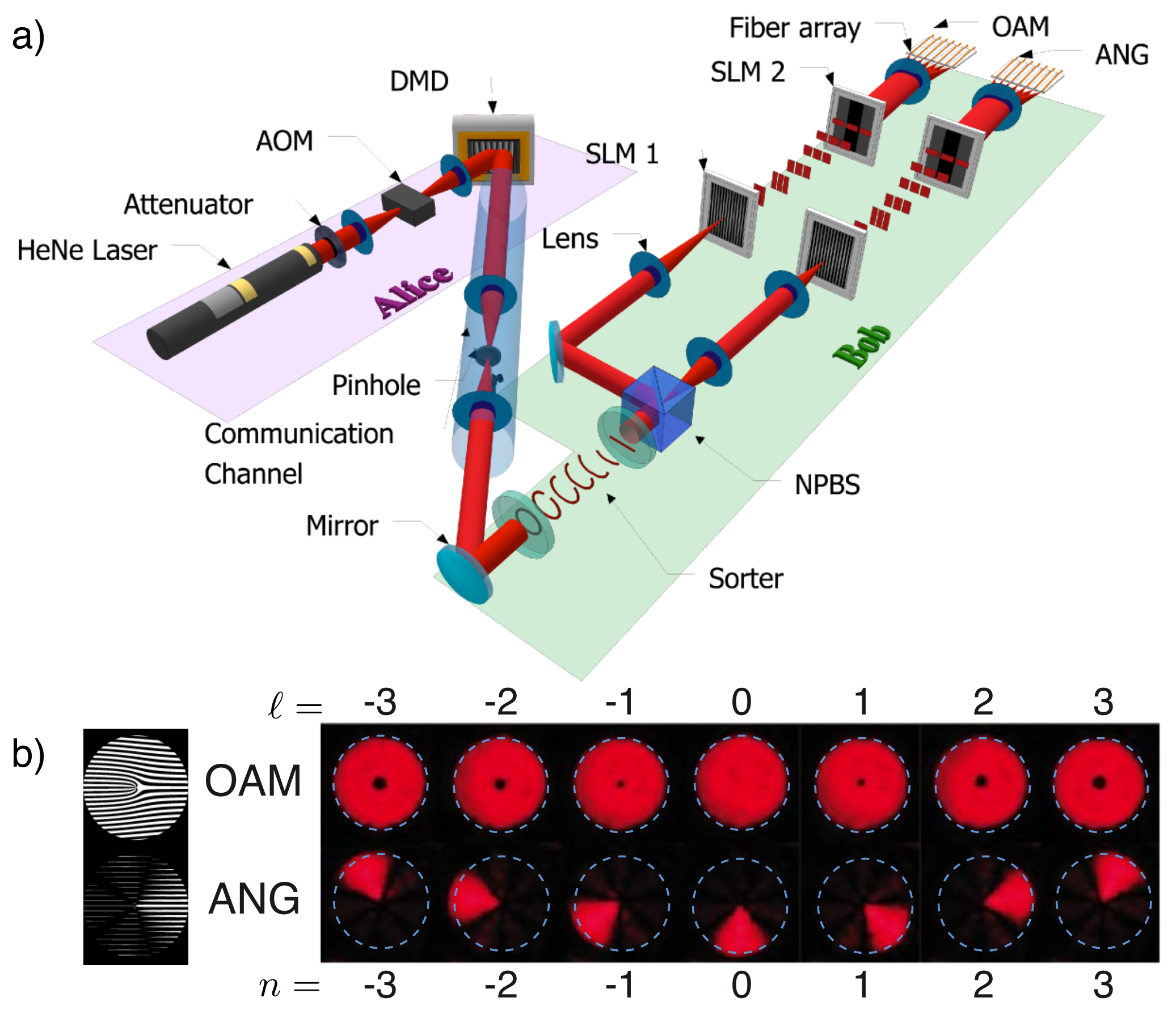}}
\caption{ \textbf{a) The experimental setup.} Alice prepares the modes by carving out pulses from a highly attenuated He-Ne laser using an AOM. The spatial mode information is impressed on these pulses with a DMD. Bob's mode sorter and fan-out elements map the OAM modes and the ANG modes into separated spots that are collected by an array of fibers. \textbf{b) The alphabet.} CCD images of produced light field profiles in two complementary spatial bases of OAM and ANG (d = 7). The intensity profile of the modes are shown on the right. An example of binary holograms used for the generation of modes in each basis is shown on the left.}
\label{fig:Setup}
\end{figure}

In our scheme, Alice randomly chooses her photons from two complementary bases. The primary encoding basis is a set of OAM vortex modes. These modes have a top-hat intensity structure and a helical phase profile characterized by $\Psi_{\mathrm{OAM}}^ {\ell} = e^{i\ell \varphi}$, where $\ell \in \{-3:3\}$ (See Fig.~\ref{fig:Setup}). We construct a mutually unbiased basis set using a linear combination of OAM modes of index $\mid\ell\mid\leq 3$ with equal amplitudes 

\be \Psi_{\mathrm{ANG}}^ {n}=\frac{1}{\sqrt{d}}\sum_{\ell=-N}^N\Psi_{\mathrm{OAM}}^ {\ell}\exp{\bigg(\frac{i2\pi n\ell}{d}\bigg)}\ee
where the dimension $d= 2N+1 = 7$ in our experiment. We refer to these modes as the angular (ANG) modes due to their localized intensity patterns (See Fig.~\ref{fig:Setup}). The ANG modes form a mutually unbiased basis with respect to the OAM basis \cite{Giovannini:2013ju,DAmbrosio:2013ir,Malik:2012ka}
\begin{equation}
 \Braket{\Psi_{\mathrm{ANG}}^ {n}|\Psi_{\mathrm{OAM}}^ {\ell}} = 
  \begin{array}{l l}
    1/d & \quad \forall \{n,\ell\}.
  \end{array} 
\end{equation}
Consequently, the measurement of a photon in the ANG basis provides no information about its OAM state and vice versa. Additionally, the ANG and the OAM bases remain mutually unbiased upon propagation (See Appendix A).

Our experimental setup is depicted in Fig.~\ref{fig:Setup}. A collimated beam from a helium-neon laser illuminates a binary hologram realized on a DMD to generate spatial modes \cite{Mirhosseini:2013go}.  We use a prepare-and-measure scheme similar to the BB84 protocol \cite{Bennett1984}. Alice initially picks a random sequence of desired OAM and ANG modes and loads the binary holograms for generating each of the modes in the sequence to the DMD's internal memory. To transmit each symbol, Alice then triggers the DMD and modulates the beam using an acousto-optic modulator to create rectangular pulses of $125$ ns width. The beam power is attenuated such that each pulse contains on average $\mu =0.1$ photons. The prepared states are then imaged to Bob's receiving aperture via a \emph{4f} telescope that forms a lossless 2-m-long communication link. 

Bob's mode sorter consists of two refractive elements for performing a log-polar to cartesian transformation \cite{Lavery2012}. Going through these elements, an OAM mode is converted to a plane wave with a tilt that is proportional to the OAM mode index $\ell$. A single lens focuses such a plane wave into a spot that is shifted by an amount proportional to $\ell$. Similarly, an ANG mode transforms to a localized spot shifted by an amount proportional to the angular index $n$ after the transformation. A beam splitter is used to randomly choose between the OAM and ANG bases. In order to reduce the overlap between the neighboring transformed modes, we use a coherent beam combination technique \cite{Mirhosseini:2013em}. A fan-out element is realized on a phase-only spatial light modulator (SLM) to divide the incident light into multiple copies. The copies are phase corrected using a second SLM and are recombined by a lens to a series of much narrower spots with the same initial spacing between them. In addition, two cylindrical lenses are realized on the SLMs to adjust the aspect ratio of the transformed modes.

\begin{figure}[t]
\centering\includegraphics[width = 0.7\textwidth]{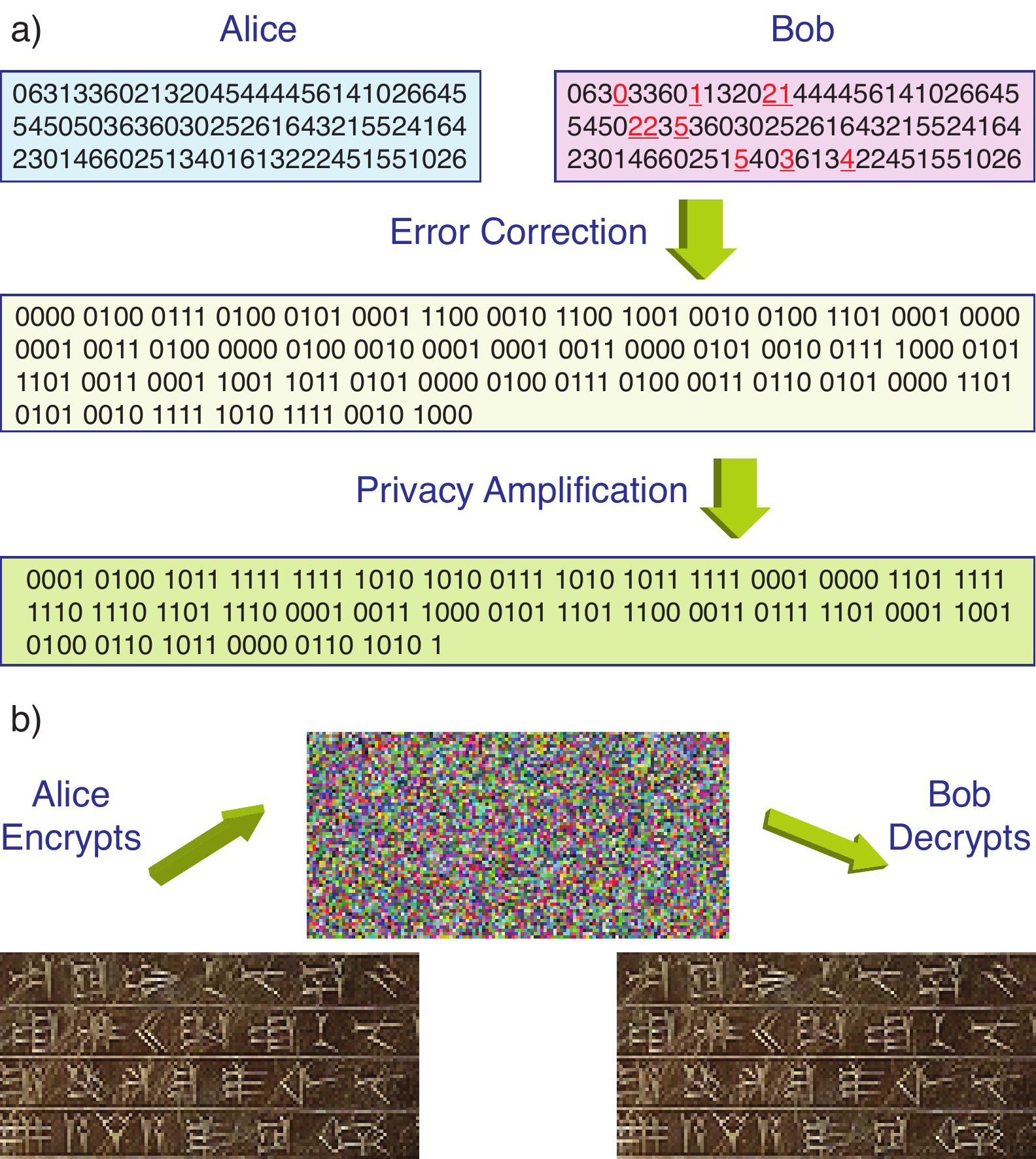}
  \caption{\textbf{Key generation} a) An example of a random sifted key from the experiment. The spatial modes are mapped to number between 0 to 6 (errors are marked in red and underlined). Each symbol is converted into a 3 digit binary number first and the binary key is randomized before the error-correction. Privacy amplification minimizes Eve's information by shortening the key length.
   b) Alice encrypts the secret message (a picture \cite{Xerxes}) using the shared secure key and Bob decrypts it.
   In this case a short key is repeated many times to match the bit length of the image. In practice, multiple use of a single key cannot provide security and a long key needs to be used.}
  \label{Key}
\end{figure}

The sorted modes are coupled to an array of fibers that are connected to avalanche photodiodes (APDs). Due to the limited number of available APDs (four), Bob's data for each group of the the symbols is taken separately and is combined later. The signal from the APDs is processed by a field-programmable gate array (FPGA) to produce photon counts using a gating signal shared with Alice. The photon detection events are finally saved in Bob's computer. Alice and Bob are also connected via a classical link realized by an ethernet cable running a TCP/IP protocol. After Alice and Bob collect a sufficiently large number of symbols, they stop the measurement. At this point, they publicly broadcast the bases used for preparation and measurement of each photon via the classical link. Alice and Bob then discard the measurements that were done in different bases. The key generated at this stage is referred to as the sifted key.

Figure~\ref{Key} shows a portion of the sifted key generated in the experiment. At this stage, the two copies of the key owned by Alice and Bob are almost identical but they contain some discrepancies. The discrepancies are due to the imperfections in our system, namely imperfect mode-sorting, detector dark counts, and ambient light. However, any errors in the key must be attributed to an eavesdropper and corrected for using the procedure of error correction. The keys are transformed to a binary form on a symbol-by-symbol basis and randomized by means of a random-number generator shared by Alice and Bob. The cascade protocol is run by Alice and Bob to fix the discrepancies by comparing the two keys in a block-by-block fashion and performing parity checks \cite{Brassard:1994wr}. After error-correction, Alice and Bob share two identical copies of the key. However, a portion of the key can still be known to Eve from eavesdropping on the quantum channel or from accessing the publicly available information transmitted in the classical channel during the error correction. Alice and Bob perform privacy amplification to minimize Eve's information. This is done by using a universal hash function for mapping the error-corrected key to a shorter random key of desired length \cite{Bennett:1995vt}. The amount by which the key is shortened at this stage is chosen by estimating the amount of information owned by Eve. Finally, the secure key is used to securely transmit information (a picture in this case) over the classical channel (See Fig.~\ref{Key}).

\begin{figure}[t]
\centerline{\includegraphics[width = 0.6\textwidth]{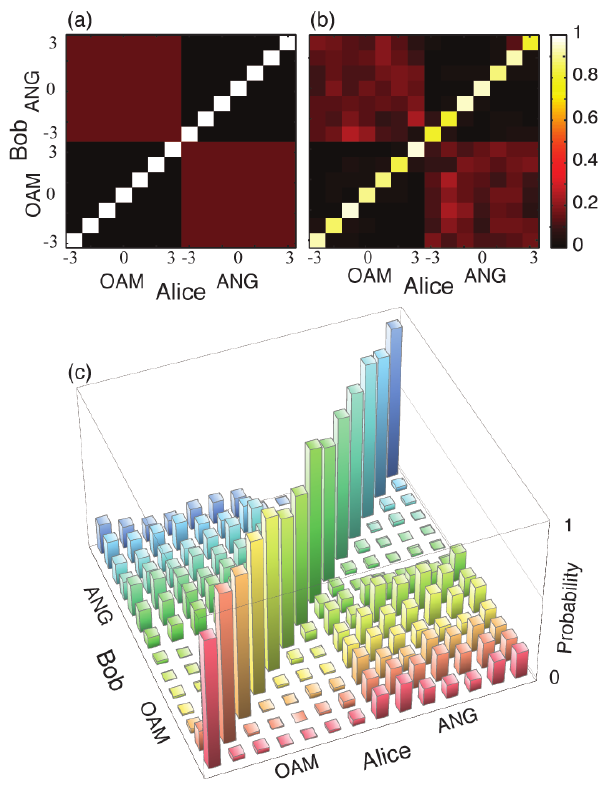}}
\caption{\textbf{Conditional probability of detection.} a) Theoretical predictions for an ideal system. The bases chosen by Alice and Bob are marked on the horizontal axis. b)The experimental results. To construct the matrix a total number of 14 million pulses with $\mu = 0.1$ photon per each are sent by Alice. The signals from APDs and the gate are collected using a 300 MHz oscilloscope on the fast acquisition mode and processed on a computer. c) 3D view of the experimental data.}
\label{fig:SMat}
\end{figure}

\section{Security Analysis}
\subsubsection*{Transmission and Detection Efficiency}

The power transmission efficiency of the mode sorter's refractive elements is measured to be $85\%$. Two Holoeye PLUTO phase-only SLMs are used for realizing the holograms for the fan-out element and its corresponding phase-correcting element. In addition, two cylindrical lenses are realized on the SLMs to adjust the aspect ratio of the transformed beams. The diffraction efficiency of each SLM is measured to be approximately $45 \%$.

The fiber array consists of 7 multi-mode fibers with 62.5 $\mu m$/125 $\mu m$ core/cladding diameters and NA = 0.275. We have measured the efficiency of coupling the transformed modes to the fiber array to be approximately $18\%$.
The detection is performed with Perkin-Elmer SPCM-AQRH-14 APDs. The quantum efficiency of the APDs is $\eta = 0.65 $.
The transmission efficiency of the optical link can be calculated as
\begin{equation}
T_{\mathrm{link}} = 0.85\times{(0.45)}^2\times0.18= 0.031.
\end{equation}
Each detectors has a typical dark-count rate of $50~c/s$ and an after-pulsing probability of $0.3\%$

\subsubsection*{Classical Information Capacity} Figure~\ref{fig:SMat} shows the conditional probability of Bob detecting each mode as a function of the mode sent by Alice. Theoretical values for the case of a system with no errors is shown on the left for comparison. In order to measure the error rate more precisely, each row of the matrix is constructed by sending the same symbol many times and detecting all possible outcomes. This scheme eliminates the time consuming procedure of switching among different spatial modes and permits a much more accurate measurement of the error-rate using a large set of sent and received symbols.

The mutual information between the sent and received symbols is defined as

\begin{equation}\label{mutualinf}
I_{AB} = \sum_{i,j} P(y_j,x_i) \log_2\left[\frac{P(y_j,x_i)}{P(x_i)P(y_j)} \right].
\label{eq:MI}
\end{equation}
Here, \(x_i\) is the event of sending symbol \(i\) and \(y_j\) is the event of detecting symbol \(j\). Assuming a uniform probability for sending \(N\) modes and a uniform probability of an error occurring in the detection, the Eq. \ref{mutualinf} can be simplified to \cite{Cerf:2002fp}
\begin{equation}
I_{AB} = \log_2 (d)  + F \log_2 (F) + (1-F) \log_2 \left(\frac{1-F}{d-1}\right).
\end{equation}

Here, $F = 1-\delta-\epsilon$ is the probability of correctly detecting each mode. The error rate caused by imperfect sorting is denoted by $\delta$ and the error rate due to the dark counts is shown by $\epsilon$. For the case of our experiment, $\delta = 0.065$ and $\epsilon = 0.04$. The total error rate $1-F$ is directly measured from data as $10.5\%$. We have identified the error from the imperfect sorting, $\delta$, by repeating the experiment in the high-light level and analyzing the results after background subtraction. The error probability $\epsilon$ can be divided to $2\%$ from stray light, $1.9\%$ from thermally-induced dark counts, and $0.15\%$ from after-pulsing. 

Using these numbers, we calculate $I_{AB}$ as 2.05 bits per photon. For comparison, the ideal value of $I_{AB}$ for $d=7$ modes is equal to $\log_2 (d) = 2.8$ bits per photon. Further, the sorting mechanism by itself would result in a value of $2.29$ bits per photon in the absence of any background or dark counts. The probability of correctly detecting each mode in the absence of dark counts, also known as the separation efficiency of the mode sorter, is calculated to be slightly more than $93\%$.

Note that the calculation of mutual information only requires the knowledge of the cross-talk and dark count values. The loss in the optical system reduces the key generation rate of the protocol. However, a uniform loss in the transmission and detection does not change the mutual information between Alice and Bob's sifted keys since the time frames with no photon-detection event are removed in the basis reconciliation procedure.

\subsubsection*{Secure key rate}
We consider a cloning-based individual attack. In this situation, the mutual information between the symbols owned by Alice and Eve can be written as \cite{Cerf:2002fp}
\begin{equation}
I_{AE} = \log_2 (d)  + F \log_2 (F_E) + (1-F_E) \log_2 \left(\frac{1-F_E}{d-1}\right).
\end{equation}
Here, $F_E$ is the fidelity of Eve's cloning machine. The value of Eve's fidelity can be optimized to gain maximum information for any given error rate detected by Bob. This quantity is shown to be \cite{Cerf:2002fp}
\begin{equation}
F_E = \frac{F}{d} + \frac{(d-1)(1-F)}{d} + \frac{2}{d}\sqrt{(d-1)F(1-F)}.
\end{equation}
For our experiment, $F_E = 0.43$, resulting in $I_{AE} = 0.35$ bits. For a sufficiently long ensemble of symbols owned by Alice, Bob, and Eve the secure key $R$ is found to be \cite{Csiszar:1978eq,Gisin:2002gb}
\begin{equation}
R_{\mathrm{net}} = R_{\mathrm{sift}} [I_{AB} - \max(I_{AE})],
\end{equation}
where $R_{\mathrm{sift}}$ is the sifted key rate. This rate can be written as a function of the pulse repetition rate $f_{\mathrm{rep}}$
\begin{equation}
R_{\mathrm{sift}} = \frac{1}{2} f_{\mathrm{rep}}\mu T_{\mathrm{link}}\eta
\end{equation}
For our experiment we have $\mu = 0.1$. The DMD used in our system (Texas Instrument DLP3000) is capable of switching between the computer holograms stored in its memory at the speed of 4kHz. We refer to this mode of operation as burst mode. Burst mode can be used to transmit a short key at $f_{\mathrm{rep}} = 4 $ kHz. The internal memory can only store up to 100 patterns, which limits the applicability of the burst mode for a long key. To generate a long random key, we instead load each hologram onto the DMD using a computer. This task along with the time needed for synchronizing the two computers reduces the raw key generation rate to approximately $f_{\mathrm{rep}} = 1 $ Hz. Using $f_{\mathrm{rep}}= 4$ kHz for the burst mode, we calculate $R_{\mathrm{sift}} = 4$ (photons/sec) and $R_{\mathrm{net}} = 6.8  $ (bits/sec). 

%
%
%

\begin{figure}[t]
\centering\includegraphics[width = 0.7\textwidth]{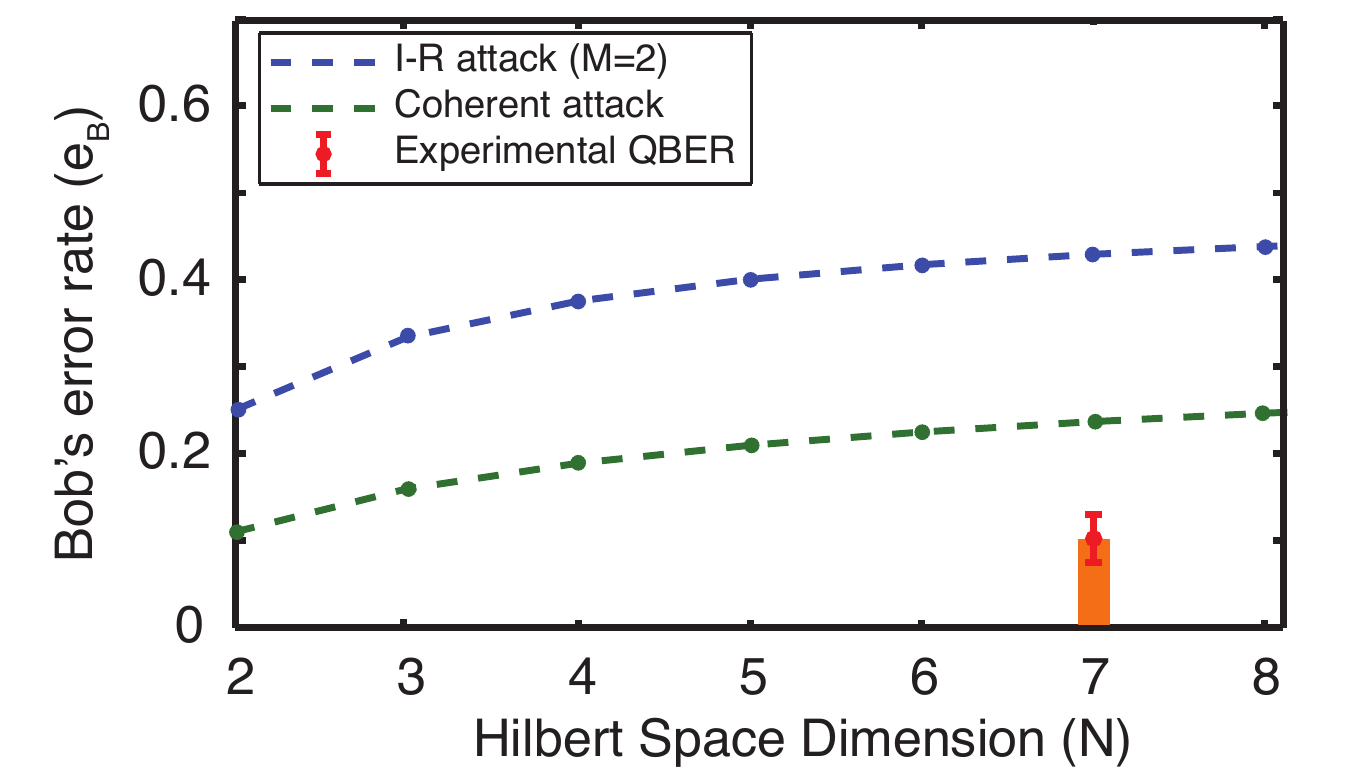}
  \caption{\textbf{Error-bound for security.} Bob's error bound calculated from theory as a function of the dimension of the Hilbert space for intercept-resend attacks with two MUBs (M=2) and coherent attacks is plotted along with QBER measured from our experimental data (the error bar shows one standard deviation). It is seen that the QBER from our experiment lies well below the theoretical bounds and hence it is sufficient for proof of security.}\label{coherent}
\end{figure}

To assess quantitatively the security of our system, we plot the values for the error bound of an intercept-resend eavesdropping attack, a coherent attack, as well as the error rate from our experimental data in Fig.~\ref{coherent}. In an intercept-resend attack, the eavesdropper (Eve) measures the state of the intercepted photon in an arbitrarily selected basis and then resends a photon prepared to be in this same state. In a coherent attack, Eve coherently probes a finite number of qudits in order to gain information about the key \cite{Cerf:2002fp}. It is evident from this graph that our experimental error rate is well below the required bounds for security against both intercept-resend and coherent attacks (Fig\,\ref{coherent}). Unlike intercept-resend attacks, the error rate for coherent attacks only depends on system dimension $d$ and is independent of the number of MUBs $M$ \cite{Cerf:2002fp}. Hence, increasing the number of MUBs beyond 2 results in a drop in the key rate without any gain in security against coherent attacks. Nevertheless, there is a clear increase in the allowed error rate for larger system dimensions, demonstrating an advantage for using a high-dimensional encoding scheme for QKD such as that of OAM. 

Using the error rate, we estimate the information gained by Eve to be 0.35 bits per sifted photon for cloning-based individual attacks. It has been argued that cloning-based eavesdropping is the optimal strategy for an individual attack on systems based on qudits \cite{Cerf:2002fp}. The portion of the key obtained by Eve is removed in the privacy amplification process, leading to a mutual information of 1.7 secure bits per sifted photon in the final key. It should be emphasized that our secure key rate analysis assumes an infinite key, while our experimental realization produces only a finite key \cite{Scarani:2008wy}. We are working toward a finite key analysis of the security, which will be reported in the future.  

In contrast to phase-only spatial light modulators (SLMs) that are limited to a frame rate of 60 Hz, the DMD used in our setup has the ability to rapidly switch between modes at a speed of 4 kHz. This is especially important considering the fact that the speed of generation of spatial modes limits the key generation rate of the system. We measure our raw key generation rate to be 16.4 bits per second (for operation in burst mode). After performing basis reconciliation and privacy amplification for a sufficiently long key, we estimate the secure key rate to be 6.5 bits per second, which is more than 3 orders of magnitude larger than that achieved in previous spatial-mode based protocols \cite{Etcheverry:2013fu}.

\subsubsection*{Photon number splitting attacks} The analysis above considers intercept-resend and cloning based attacks. The photon-number-splitting (PNS) attack is yet another possibility when the protocol uses imperfect single-photon sources. In our protocol, weak coherent pulses (WCPs) are used for realizing approximate single photon pulses. For small values of photon-number expectation values, a WCP contains either zero or one photon most of the times. However, the pulse may still contain multiple photons with a non-zero probability. In this situation, Eve can split one photon off a multi-photon signal, without disturbing the polarization or the spatial structure of the signal \cite{Yuen:1996ji,Huttner:1995ck, Lutkenhaus:2000fn}. Eve can measure her photon after basis reconciliation and obtain complete information without causing any error in the signal. 

The PNS attacks pose a more serious threat in the presence of loss. In this situation, Eve can, in principle, replace the lossy channel with a perfect quantum channel and only transmit to Bob the signals of her choice. An inequality has been developed in \cite{Brassard:2000dc} that serves as a necessary condition for security
\begin{equation}
p_{\mathrm{detection}} > p_{\mathrm{multi}}.
\end{equation}
Here, $p_{\mathrm{detection}} $ is the total probability of detection events for Bob, and $p_{\mathrm{multi}}$ is the probability of having a pulse with multi-photons. For a coherent state with an average photon number o f $\mu$, we have $P(n,\mu) = \frac{\mu^n}{n!}e^{-\mu}$, and consequently

\begin{equation}
p_{\mathrm{multi}} = \sum_{n=2}^{\infty} P(n,\mu) \approx \frac{\mu}{2}.
\end{equation}

The total probability of detection events can be calculated by considering the signal detection events and the detector dark counts

\begin{equation}
 p_{\mathrm{detection}} = p_{\mathrm{signal}} + p_{\mathrm{dark}} - p_{\mathrm{signal}}p_{\mathrm{dark}} \approx p_{\mathrm{signal}} + p_{\mathrm{dark}}.
\end{equation}
Using the parameters for our experiments we have $p_{\mathrm{multi}} = 5\times 10^{-3}$ and $ p_{\mathrm{detection}} = \mu T_{\mathrm{link}}\eta + p_{\mathrm{dark}}  \simeq 2\times 10^{-3}$. It can be see that $p_{\mathrm{detection}} < p_{\mathrm{multi}}$ and hence security against PNS attacks cannot be guaranteed in our current implementation. We discuss the possible approaches for providing security against PNS attacks in the following section. 

\section{Steps towards practical QKD with OAM modes}

The secure key rate of a QKD system is considered as the ultimate metric for quantifying its performance. In the above sections we described how a high-dimensional encoding scheme such as that of OAM  can potentially benefit the secure key rate of QKD protocols. While our experiment provides significant improvements in realizing OAM-based QKD, several challenges need to be addressed before such a protocol can be employed for practical applications. In addition to a high secure key rate, a free-space QKD link needs to provide reliable long distance operation, efficient classical post-processing, ease of alignment, and a rigorous security analysis considering the finite size of practical keys. While some of these requirements can be achieved by increasing the degree of sophistication of the system or through the use of state-of-the-art technology, there is still need for theoretical and experimental research before a full demonstration can be performed. Below, we address some of the current experimental limitations.

\subsubsection*{Fast key generation:}The secure key rate scales proportionally with the raw key generation rate. A fast rate can be achieved by using an array of static holograms for generating multiple modes and then multiplexing the modes using a series of beam splitters. A cheaper and much simpler alternative to this approach is to modulate the intensity and the phase of a laser beam in real time. Although holography techniques can achieve kHz mode switching rates using DMDs \cite{Mirhosseini:2013go}, practical QKD would require key rates in the GHz regime. OAM modes can be generated and switched at MHz rates using on-chip resonators \cite{Strain:2014gh} and potentially at GHz rates using Q-plates \cite{Karimi:2009ge}. However, to generate the states in the ANG basis one would need to modulate both the amplitude and phase of the beam, a task that best suits free-space holography.  Recently, a method involving static holograms realized on an SLM and an AOM for switching between them achieved a MHz mode-switching rate \cite{Radwell:2014fy}. Unfortunately, the change in the wavelength caused by the AOM forbids applying this method for QKS since side-channel attacks could be performed based on spectrum. More research is required for developing high-speed methods for arbitrary generation of free-space spatial modes.
 
\subsubsection*{High throughput:}The key generation rate drops as loss increases. Moreover, high loss in the communication link makes the protocol vulnerable to PNS attacks. The throughput of our detection system can be readily increased by employing high efficiency spatial light modulators, or AR-coated custom refractive elements. This would translate to a six-fold increase in the transmission efficiency in our experiment. Additionally, the amount of loss due to the scattering in the air can be minimized by operating in the near-infrared regime.

\subsubsection*{Turbulence mitigation:} Atmospheric Turbulence results in degradation of the spatial profile of the modes upon propagation. This results in mixing of the neighboring OAM and ANG modes. The effects of turbulence on OAM modes has been a topic of extensive studies \cite{Malik:2012ka,Tyler2009,Paterson:2005km}. Common solutions for mitigating the adverse effects of atmospheric turbulence include the use of every other mode for encoding \cite{rodenburg2012influence} and utilization of adaptive optics systems \cite{Rodenburg:2014dn}. Recently, long-haul free-space propagation of OAM modes has been realized using novel detection schemes \cite{Krenn:2014wj}.

\subsubsection*{Larger dimensionality:} Increasing the number of the modes increases the information carried by each single photon and results in a higher secure bit rate. Previously, we have shown that our mode sorter is capable of sorting 25 OAM and ANG modes with an average mutual information of 4.17 bits per detected photons \cite{Mirhosseini:2013em}. Consequently, the encoded information per photon can be readily increased by increasing the number of APDs in the experiment. Ultimately, the number of modes supported by the optical link is limited by the sizes of the transmitting and receiving apertures.
 
\subsubsection*{Universal security proof:} A comprehensive security proof for a QKD system should include both the fundamental and practical properties of the physical system. The security analysis presented in this work assumes an infinitely long key. For a finite key, the efficiency of classical post-processing need to be measured and used for a more rigorous calculation of the secure key rate \cite{Scarani:2009zz}. In addition, our analysis assumes a uniform error-rate and transmission efficiency for all the modes. While our data demonstrates the rationale for this approximation, a more comprehensive analysis needs to include the effects of non-uniform error rate and loss in a multi-level QKD system. More research is needed to establish the theoretical framework for security analysis of such systems.

Finally, we have demonstrated security against intercept-resend and cloning-based attacks, while our implementation remains vulnerable to photon-number-splitting attacks. This limitation can be avoided by either reducing the loss in the system or by employing the decoy-state protocol \cite{Hwang:2003wm}

\section{Conclusions}

In conclusion, we have demonstrated that practical QKD based on spatial-mode encoding is realizable with current technology. We perform a proof-of-principle experiment that uses a 7-dimensional alphabet encoded in OAM and in the mutual unbiased basis of ANG modes. We implement a fast mode-generation technique that uses a DMD to generate spatial modes at a speed of 4 kHz and a mode-transformation technique that is able to measure the OAM and ANG content of a photon with an accuracy of 93\%. Using these two methods, we achieve a mutual information of 2.05 bits per sifted photon, which is more than twice the maximum allowable capacity of a two-dimensional QKD system. The QBER of our scheme is measured to be 10.5$\%$, which is sufficient for proving unconditional security against coherent and individual eavesdropping attacks for an infinite key. In addition, we lay out a clear path for how our system can be enhanced to perform practical, high-dimensional QKD using current technology. For example, our QBER can be significantly reduced through the use of better detectors and a more sophisticated mode-transformation technique. Our experiment opens the door to realizing real-world, multi-level quantum communication systems with record capacities and levels of security.\\

\bibliographystyle{njp.bst}
\bibliography{QKDRef,SuppInf}{}

\appendix

\section{Diffraction and radial modes}
It is known that OAM modes with larger mode index $\ell$ have a larger rms (root mean squared) radius in the far field \cite{Phillips:1983gm}. Given the different radial profiles of different OAM modes, one might be concerned that the security of the protocol is compromised. More specifically, we are interested to know whether Eve can gain information by using the photons that fall outside Bob's aperture? The details of the diffraction behavior depends on the radial profile of the OAM modes as defined in the transmitter's aperture. We consider OAM modes with a top-hat intensity profile such as the ones used in our experiment. Since diffraction plays no role once the Fresnel number is sufficiently large, this issue does not directly apply to our experiment. However, the answer to the question for a low-Fresnel-number system is not trivial. Below, we show that it is impossible for Eve to gain any information by performing this attack regardless of the Fresnel number of the system. 

\subsubsection*{Intercept and intercept attack}
In an intercept and intercept attack (also known as denial of service) Eve detects a photon without retransmitting it. Since Alice and Bob disregards the frames with no photons in basis reconciliation, Eve will not have any mutual information with the sifted key and hence there is no loss of security.
\subsubsection*{Intercept and resend attack}
\begin{figure}[t]
\centering\includegraphics[scale = 0.7]{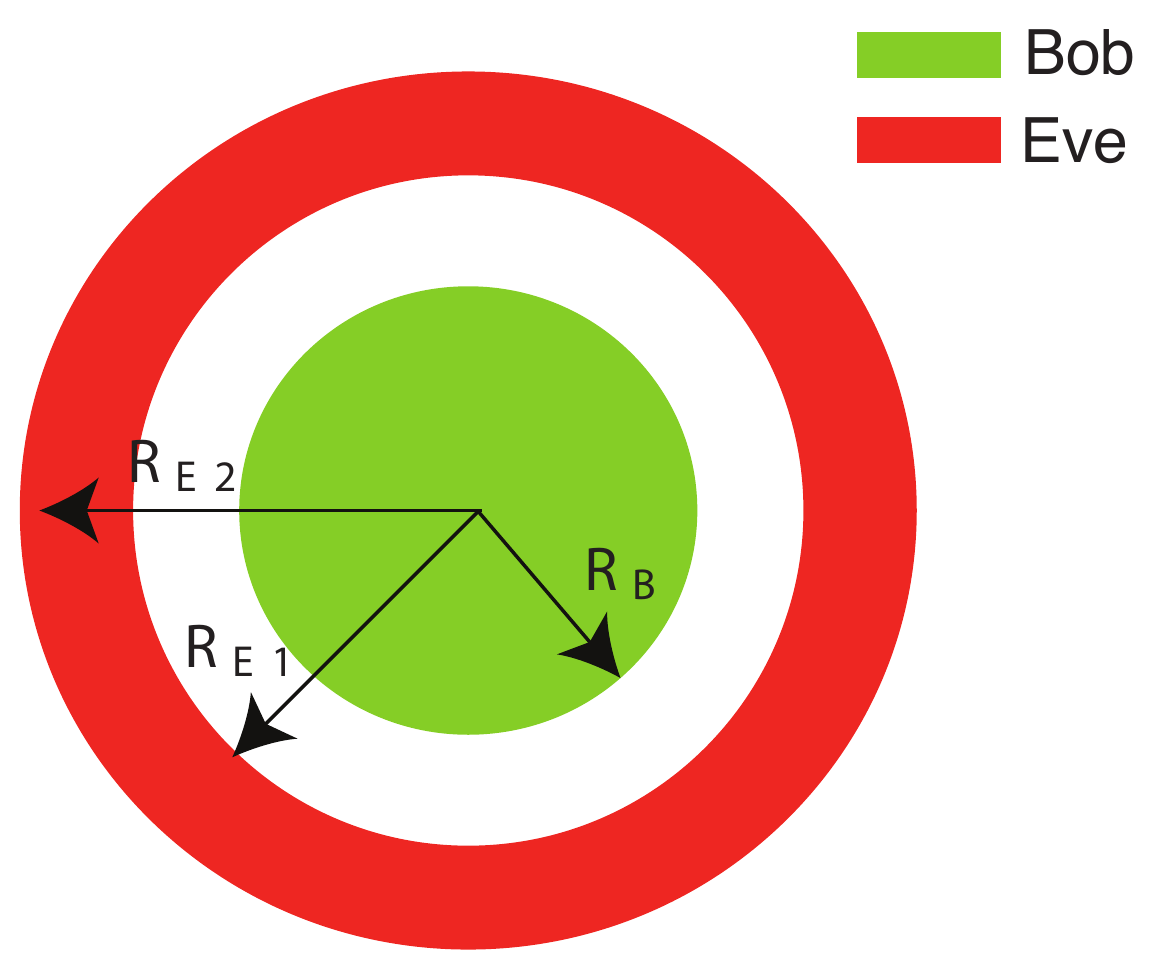}
  \caption{\textbf{Bob's and Eve's receiving apertures} }
  \label{Ap}
\end{figure}
In this type of attack Eve detects a photon that falls outside of Bob's apertures in a random basis. She then resends a photon in a state determined by the result of her measurement into the Bob's aperture. The geometry for Bob's and Eve's apertures are shown in Fig. \ref{Ap}. Note that the $R_B, R_{E1}$ and $R_{E2}$ are all bound between zero and infinity and so we are considering the most general case.

We demonstrate an OAM mode's optical field at any plane $z$ by $\Psi_{\mathrm{OAM}}^ {\ell} (z)$. The probability for detecting \textbf{each} of the OAM modes by Eve can be written as
\begin{equation}
  P_{\mathrm{OAM}}^ {\ell} (z)= \int_{R_{E1}}^{R_{E2}}  \left|\Psi_{\mathrm{OAM}}^ {\ell} (z)\right|^2 r dr.
\end{equation}
Assuming Alice chooses to send OAM modes randomly with a probability of $1/d$ (where $d =2N+1$ is the total number of modes used), the probability for detecting \textbf{any} OAM mode will be 
\begin{equation}
  P_{\mathrm{OAM}}(z)= \frac{1}{{d}}\sum_{\ell=-N}^{N} \int_{R_{E1}}^{R_{E2}}  \left|\Psi_{\mathrm{OAM}}^ {\ell} (z)\right|^2 r dr,
\end{equation}
Similarly, the probability for detecting \textbf{each} of the ANG modes can be written as 
\begin{equation}
 P_{\mathrm{ANG}}^ {n} (z)= \int_{R_{E1}}^{R_{E2}}  \left| \Psi_{\mathrm{ANG}}^ {n} (z) \right|^2 r dr, 
\end{equation} 
and the probability for detecting \textbf{any} ANG mode will be
\begin{equation}
 P_{\mathrm{ANG}} (z)= \frac{1}{{d}}\sum_{n=-N}^{N} \int_{R_{E1}}^{R_{E2}}  \left|\Psi_{\mathrm{ANG}}^ {n} (z)\right|^2 r dr, 
 \label{ANG}
\end{equation} 
We can simplify Eq. \ref{ANG} using the definition of the ANG modes
 \begin{equation}
  \Psi_{\mathrm{ANG}}^ {n}=\frac{1}{\sqrt{d}}\sum_{\ell=-N}^N\Psi_{\mathrm{OAM}}^ {\ell}\exp{\bigg(\frac{i2\pi n\ell}{d}\bigg)}.
 \end{equation}
Substituting $ \Psi_{\mathrm{ANG}}^ {n}$ we get
\begin{eqnarray}
 P_{\mathrm{ANG}} (z) &= & \frac{1}{{d}}\sum_{n= - N}^N \int_{R_{E1}}^{R_{E2}}  \left|\frac{1}{\sqrt{d}}\sum_{\ell=-N}^N\Psi_{\mathrm{OAM}}^ {\ell}\exp{\bigg(\frac{i2\pi n\ell}{d}\bigg)}\right|^2 r dr \nonumber\\
 & = & \frac{1}{{d}}\int_{R_{E1}}^{R_{E2}}\sum_{n= - N}^N \sum_{\ell=-N}^N\sum_{\ell'=-N}^N \frac{1}{{d}} \Psi_{\mathrm{OAM}}^ {\ell}\Psi_{\mathrm{OAM}}^ {*\ell'}\exp{\bigg[\frac{i2\pi n(\ell-\ell')}{d}\bigg]} r dr\nonumber\\
 & = & \frac{1}{{d}}\int_{R_{E1}}^{R_{E2}}\sum_{\ell=-N}^N\sum_{\ell'=-N}^N \Psi_{\mathrm{OAM}}^ {\ell}\Psi_{\mathrm{OAM}}^ {*\ell'}\sum_{n= - N}^N \frac{1}{{d}} \exp{\bigg[\frac{i2\pi n(\ell-\ell')}{d}\bigg]} r dr\nonumber\\
  & = & \frac{1}{{d}}\int_{R_{E1}}^{R_{E2}}\sum_{\ell=-N}^N\sum_{\ell'=-N}^N \Psi_{\mathrm{OAM}}^ {\ell}\Psi_{\mathrm{OAM}}^ {*\ell'} \delta_{\ell,\ell'} r dr\nonumber\\
  & = & \frac{1}{{d}}\int_{R_{E1}}^{R_{E2}}\sum_{\ell=-N}^N \left|\Psi_{\mathrm{OAM}}^ {\ell}\right|^2 r dr\nonumber\\
  & = & P_{\mathrm{OAM}}(z).
\end{eqnarray} 
The above result shows that it is equally probable for Eve to detect a photon from either the OAM basis or the ANG basis using the beam from any range of radii. Consequently, it is impossible for Eve to gain information by detecting a photon from outside of Bob's aperture. Moreover, the result remains valid at any plane $z$ from near field to far field.

\end{document}